\documentclass[conference]{IEEEtran}
\IEEEoverridecommandlockouts
\usepackage{algorithm}
\usepackage{algorithmic}
\usepackage{booktabs}
\usepackage{cite}
\usepackage{subcaption}
\usepackage{amsmath,amssymb,amsfonts}
\usepackage{graphicx}
\usepackage{textcomp}
\usepackage{hyperref} 
\usepackage{xcolor}
\def\BibTeX{{\rm B\kern-.05em{\sc i\kern-.025em b}\kern-.08em
    T\kern-.1667em\lower.7ex\hbox{E}\kern-.125emX}}
    
\title{RFI Detection and Identification at OVRO Using Pseudonymetry}

\author{
\IEEEauthorblockN{Meles G. Weldegebriel}
\IEEEauthorblockA{
Electrical and Systems Engineering\\
Washington University in St. Louis\\
St. Louis, MO, USA}

\and
\IEEEauthorblockN{Zihan Li}
\IEEEauthorblockA{
Computer Science and Engineering\\
Washington University in St. Louis\\
St. Louis, MO, USA}

\and[\hfill\mbox{}\par\mbox{}\hfill]
\IEEEauthorblockN{Greg Hellbourg}
\IEEEauthorblockA{
Cahill Center for Astronomy \& Astrophysics\\
California Institute of Technology\\
Pasadena, CA, USA}

\and
\IEEEauthorblockN{Ning Zhang}
\IEEEauthorblockA{
Computer Science and Engineering\\
Washington University in St. Louis\\
St. Louis, MO, USA}

\and
\IEEEauthorblockN{Neal Patwari}
\IEEEauthorblockA{
Price College of Engineering\\
University of Utah\\
Salt Lake City, UT, USA}
}

\begin{document}
\maketitle

\begin{abstract}
    Protecting passive radio astronomy observatories from unintended radio-frequency interference (RFI) is increasingly challenging as wireless activity expands near protected bands. While radio quiet zones, database-driven coordination, and post-processing mitigation can reduce interference risk, they often lack the ability to \emph{attribute} detected RFI to a specific transmitter—particularly in low signal-to-noise ratio (SNR) regimes where conventional demodulation is infeasible. This paper presents the first over-the-air field demonstration of \textit{Pseudonymetry} at the Owens Valley Radio Observatory (OVRO), evaluating an accountable coexistence approach between heterogeneous systems: an SDR-based narrowband OFDM transmitter and a wideband radio telescope backend. The transmitter embeds a pseudonym watermark on a dedicated OFDM subcarrier using coded power modulation, while OVRO passively extracts the watermark from standard backend spectrogram (power) products without IQ access. We develop a spectrogram-only receiver that performs correlation-based packet alignment, compensates timing-resolution mismatch via resampling, and decodes pseudonym bits using energy-domain template matching. Field results across $-20$ to $+5$~dB SNR show that pseudonym watermarks can be recovered at low SNR, enabling practical transmitter attribution using only passive backend measurements. These findings suggest that observatories can support lightweight accountability mechanisms that complement dynamic protection and enforcement-oriented spectrum sharing frameworks.
\end{abstract}
\begin{IEEEkeywords}
Radio Frequency Interference, Passive Receiver, Radio Astronomy, Pseudonymetry, Interference Mitigation.
\end{IEEEkeywords}

\thispagestyle{plain}  
\pagestyle{plain} 

\section{Introduction}



Radio-frequency interference (RFI) to passive radio telescopes is a very serious and growing problem. Today wireless devices are increasing fast in number, applications, and deployment density, and therefore the spectrum becomes more congested. This makes it more likely that active wireless transmissions happen close to protected passive bands. Radio telescopes are especially sensitive receivers because they are designed to detect extremely weak cosmic signals. Due to this, even a very weak interference signal that is not important for a normal communication receiver can still corrupt the telescope observations and reduce the quality of scientific data products. Facilities such as the Owens Valley Radio Observatory (OVRO) therefore need effective methods not only to detect RFI, but also to support safe and sustainable coexistence between passive scientific receivers and active wireless systems.

There is important prior work to manage RFI in radio astronomy. Classical protection approaches include geographic radio quiet zones and propagation-model-based planning~\cite{Wilson2014propagation}. When interference happens, observatories also use reactive mitigation pipelines such as filtering, excision, adaptive flagging, and statistical post-processing~\cite{Ellingson2005rfi,Ford2014rfi,Offringa2013}. These methods are useful and they work well in many cases. However, they still leave serious limitations. For example, quiet zones can become too conservative and reduce spectrum utilization. Also, post-processing cannot always recover lost observation data, especially when the RFI is weak, intermittent, bursty, or overlapping in frequency. As spectrum sharing becomes more dynamic, relying only on avoidance and excision is not enough to protect the passive receivers and also allow efficient spectrum access.

To address these challenges, recent research has explored more flexible coexistence frameworks between active and passive services. Some works propose cooperative mitigation mechanisms~\cite{Anastasopoulos2013satellite}. Other works propose dynamic protection concepts such as Dynamic Protection Areas (DPAs) and adaptive exclusion zones~\cite{Papadopoulos2023DPA,Munira2024DynamicPZ}. Machine learning methods have also been studied for automated RFI detection and classification in telescope data~\cite{Zhang2022deepRFI}. These efforts help to make spectrum sharing more dynamic than only propagation-model-based quiet zones. However, there is still a major operational gap. Even when a telescope detects harmful RFI, it is often difficult to know \emph{which transmitter} produced it. Without device-level attribution, enforcement is weak, coordination is difficult to trigger, and dynamic protection frameworks tend to remain conservative. In practice, sustainable coexistence needs not only detection, but also practical mechanisms for \emph{accountability} and \emph{enforcement}.

This paper addresses this gap using \textit{Pseudonymetry}~\cite{Meles2022pseudonymetry}, which is a cooperative coexistence approach where a secondary transmitter embeds a lightweight identifier (pseudonym watermark) directly into its transmission. The watermark is designed to work in extremely low-SNR conditions using energy-domain processing. This makes coexistence monitoring move from only \textit{``RFI detection''} to \textit{``RFI detection with transmitter attribution.''} In the long-term vision, this type of attribution can support closed-loop interference management. For example, identification can trigger automated reporting, cooperative shutoff, and other enforcement actions, which can reduce uncertainty for dynamic protection frameworks.

A key challenge for adopting pseudonymetry at real observatories is receiver integration. Radio telescope backends are designed mainly for science data products, and they typically provide time-frequency \emph{spectrograms} (power measurements), not raw IQ samples. Because of this, many conventional watermark receivers that assume coherent processing and IQ access cannot be directly applied. This leads to a practical and important question: \emph{can pseudonym watermarks be extracted and decoded using only the spectrogram outputs that are already produced by radio telescope receivers?}

We answer this question through an over-the-air field trial at OVRO using the Deep Synoptic Array (DSA) backend spectrometer. An SDR-based OFDM transmitter operates at 920 MHz and embeds a pseudonym watermark on a single OFDM subcarrier. Importantly, the DSA backend provides spectrogram-only measurements, which matches real passive receiver constraints. We therefore develop and evaluate a \emph{spectrogram-only} pseudonym receiver software chain. Our receiver performs correlation-based packet alignment, compensates timing-resolution mismatch using resampling, and decodes pseudonym bits using energy-domain template matching. Using only standard telescope backend products, we show that pseudonym watermarks can be recovered under low-SNR conditions relevant to passive protection criteria~\cite{ITU-RA769-2}.

\textbf{Contributions.}
\begin{itemize}
    \item \textbf{Experimental evidence for spectrogram-only pseudonymetry at a radio observatory:} We present an OVRO field trial showing that transmitter-embedded pseudonym watermarks can be decoded using only telescope backend spectrograms (no IQ access).
    \item \textbf{Practical receiver software chain for passive backends:} We provide a spectrogram-only decoding pipeline that performs synchronization under unknown timing and corrects chip-resolution mismatch using correlation and resampling.
    \item \textbf{Implications for accountability-oriented coexistence:} We evaluate performance across $-20$~dB to $+5$~dB SNR and discuss how spectrogram-based attribution can complement dynamic protection and enforcement-oriented spectrum sharing frameworks.
\end{itemize}

\section{Background and Related Work}
\subsection{RFI mitigation in radio astronomy}
RFI mitigation is a very important part of radio astronomy operation. In practice, observatories use a combination of site protection rules, front-end filtering, and post-processing methods to reduce interference impact. Techniques such as adaptive thresholding, excision, and flagging are widely used \cite{Ellingson2005rfi,Ford2014rfi,Offringa2013}. For example, LOFAR demonstrated a practical mitigation pipeline with strong real-world results \cite{Offringa2013}. 

However, most of these mitigation methods are reactive. They mainly try to clean the data after the RFI already happened. Also, they do not provide device-level attribution, meaning they usually cannot tell which transmitter caused the interference. When interference becomes more frequent, reactive cleaning becomes expensive, time-consuming, and sometimes not effective. This motivates coexistence-oriented approaches that include better monitoring and enforcement capability.

\subsection{Protection criteria and low-SNR interference}
For passive services like radio astronomy, protection criteria are very strict. The ITU-R RA.769 recommendation defines harmful interference thresholds for radio astronomy and shows that even extremely weak emissions can be harmful \cite{ITU-RA769-2}. This means that interference signals that are far below the reception threshold of normal communication systems can still damage telescope observations. Therefore, we need techniques that can work in very low SNR regimes and do not depend on conventional payload demodulation.

\subsection{Coexistence and dynamic protection}
Coexistence between active and passive services has been studied in different settings. For example, some work considers interference mitigation strategies for satellite systems coexisting with passive receivers \cite{Anastasopoulos2013satellite}. More recently, dynamic protection approaches propose adapting access rules around observatories based on conditions. Dynamic Protection Area (DPA) neighborhood methods aim to manage sharing more actively and reduce overly conservative protection \cite{Papadopoulos2023DPA}. Dynamic protection zones also try to adjust spectrum access in real time \cite{Munira2024DynamicPZ}. 

These methods can help to improve spectrum utilization, but in practice they need reliable monitoring and enforcement tools. If the system cannot reliably detect and attribute interference events, then dynamic protection still tends to become conservative.

\subsection{Machine learning for RFI detection}
Machine learning approaches, including deep learning, have been proposed for RFI detection in radio astronomy data \cite{Zhang2022deepRFI}. These methods can detect complex patterns and can support automation. However, detection alone is not enough for accountability, because detection does not provide attribution. In other words, the telescope may know that RFI exists, but it still does not know which transmitter caused it. Our work is complementary: ML can help detect interference events, while pseudonymetry can provide transmitter identity when the transmitter cooperates.

\subsection{Pseudonymetry for cooperative accountability}
Pseudonymetry is a cooperative coexistence approach that provides device-level accountability by embedding a lightweight identifier (called a pseudonym watermark) directly into the transmitted waveform \cite{Meles2022pseudonymetry}. The main idea is that a transmitter can include an additional low-overhead signature that a passive receiver can decode, even when normal payload demodulation is not possible. 

A key feature of pseudonymetry is that it is designed for very low SNR operation. Instead of relying on coherent communication decoding, the pseudonym watermark can be decoded using energy-domain processing. This is important for passive services such as radio astronomy, because harmful interference can occur far below the demodulation threshold of conventional receivers \cite{ITU-RA769-2}. 

In the longer-term vision, pseudonymetry can support closed-loop coexistence. For example, if an observatory detects and attributes harmful RFI to a specific pseudonym, this information can trigger reporting, cooperative shutoff, or enforcement actions. This can reduce uncertainty and allow dynamic protection frameworks to operate in a less conservative way.

\subsection{Positioning of this work}
This paper builds on prior pseudonymetry concepts and focuses on a key missing step for practical deployment in passive services: \textbf{integration with radio telescope backends}. In real observatories, the backend is designed mainly for science products and often provides time-frequency \emph{spectrograms} (power measurements), not raw IQ samples. As a result, many conventional watermark receivers that assume IQ access cannot be directly applied.

Therefore, our work contributes a practical spectrogram-only receiver pipeline that supports synchronization, timing mismatch correction, and pseudonym decoding using standard backend outputs. This provides a realistic pathway for accountability-oriented monitoring that can complement dynamic protection and enforcement-oriented spectrum sharing frameworks.

\section{OVRO Field Trial: System Overview and Design Goals}
\subsection{Experiment motivation}
The OVRO environment is challenging and realistic. The telescope backend is designed for scientific spectrogram products, not for communication decoding. We ask a practical coexistence question: can a passive receiver perform accountability-oriented attribution using only backend spectrogram output, without adding a dedicated receiver chain?

Also, the field trial models realistic weak interference reception: the transmitter signal is received partly through DSA antenna sidelobes. This is important because harmful interference can reach passive receivers through indirect paths even when transmitters are not close or aligned.

\subsection{System overview}
The field trial includes:
\begin{itemize}
    \item \textbf{Secondary transmitter (cooperative RFI source):} USRP B210 transmitting OFDM at 920MHz with a pseudonym watermark on one subcarrier.
    \item \textbf{Passive receiver:} OVRO/DSA backend spectrometer producing time-frequency power spectrograms.
    \item \textbf{Receiver pipeline:} software post-processing for packet extraction and pseudonym decoding (attribution).
\end{itemize}

The receiver does not know transmitter start time. It only observes energy within a wide bandwidth. The RFI appears as narrowband energy inside the spectrogram.

\begin{figure}[t]
    \centering
    \includegraphics[width=\linewidth]{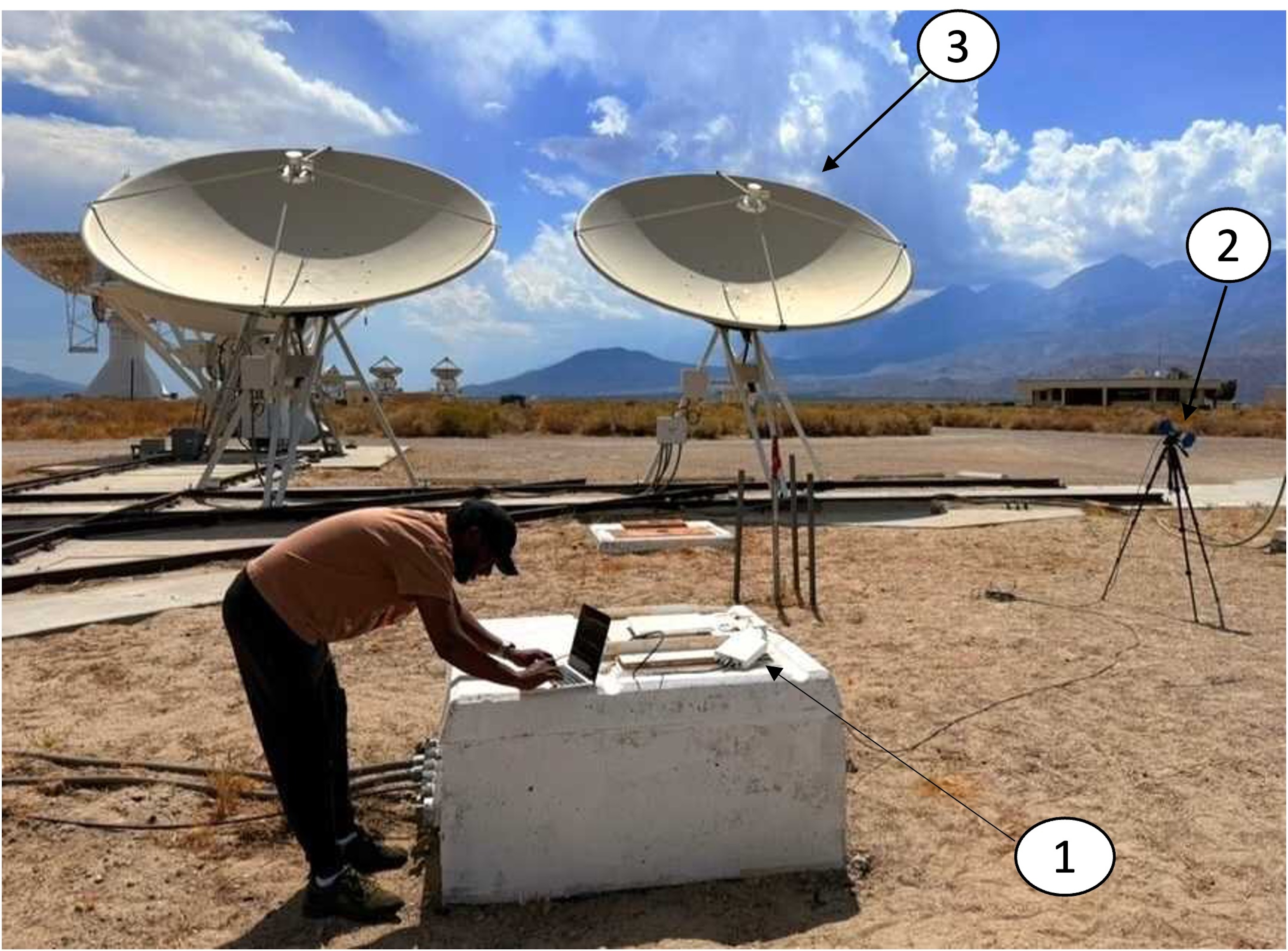}
    \caption{Field trial setup at OVRO. (1) USRP B210 with laptop control, (2) Bicolog 20300 transmit antenna, (3) DSA-110 receiving antenna.}
    \label{fig:ovro_setup}
\end{figure}

\subsection{Design goals}
\textbf{G1: Spectrogram-only attribution.} Use only power spectrogram, no IQ.

\textbf{G2: Synchronization under unknown timing.} Detect packet start by correlation.

\textbf{G3: Robustness to resolution mismatch.} Correct drift using resampling.

\textbf{G4: Low-SNR decoding.} Attribute weak interference consistent with strict passive protection criteria \cite{ITU-RA769-2}.

\subsection{Design rationale for OVRO backend}
We intentionally design the system to match backend limitations. In practice, radio astronomy backends already generate spectrograms for science. If pseudonymetry can use this same data, deployment cost is reduced. Also, spectrogram-only decoding means the approach can be applied retroactively to stored data, enabling offline attribution analysis of interference events. This supports enforcement-oriented reporting and coordination within dynamic protection frameworks.

\section{Pseudonym Watermark Design for Spectrogram-Based Identification}
We implement pseudonymetry as a code-modulated watermark on a dedicated OFDM subcarrier. The waveform uses 64-subcarrier OFDM at 6~MHz sampling rate, giving subcarrier spacing:
\begin{equation}
\Delta f = \frac{6~\text{MHz}}{64} = 93.75~\text{kHz}.
\end{equation}
Only one subcarrier carries the watermark. This concentrates energy and makes it visible in a single spectrogram frequency bin. It also limits watermark overhead to about $1/64$ of OFDM resources.

\subsection{PN-based coded modulation}
We use a maximum-length PN (m-sequence) of length 15 to generate the chip-level watermark pattern. Each pseudonym bit is represented by 15 PN chips, and the watermark is embedded using coded pulse amplitude modulation (CPAM) in the power domain. In particular, each chip is mapped to one of two power levels (e.g., $1+\alpha$ and $1-\alpha$), producing the chip-level watermark code $Q_p(t)$.

In our implementation, we use 6 samples per chip at the transmitter, so each pseudonym bit spans $15\times 6 = 90$ samples. A pseudonym packet contains 28 pseudonym bits, so one packet has $90\times 28 = 2520$ samples. Packets are repeated continuously for around 5 seconds, giving about 163 packets. This repetition improves robustness and supports low-SNR transmitter attribution.

\begin{figure}[t]
    \centering
    \includegraphics[width=\linewidth]{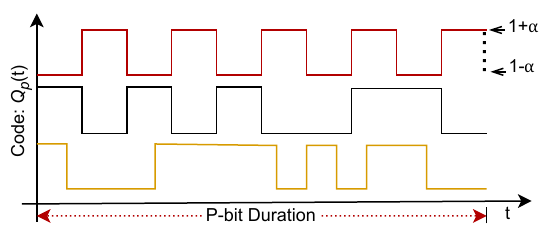}
    \caption{Example watermark code $Q_p(t)$ over one pseudonym-bit duration using CPAM, where a length-15 maximum-length PN chip pattern is mapped to two power levels (e.g., $1+\alpha$ and $1-\alpha$). Adapted from~\cite{Meles2025StopSec}.}
    \label{fig:watermark_codes}
\end{figure}

The watermark code structure shown in Fig.~\ref{fig:watermark_codes} is adapted from our prior work~\cite{Meles2025StopSec}. 

\subsection{Energy-domain watermark representation}
Chip values control power: chip ``0'' is low power and chip ``1'' is high power. Therefore, the watermark appears as a structured power variation pattern in the extracted spectrogram time series.

\begin{figure}[t]
    \centering
    \includegraphics[width=\linewidth]{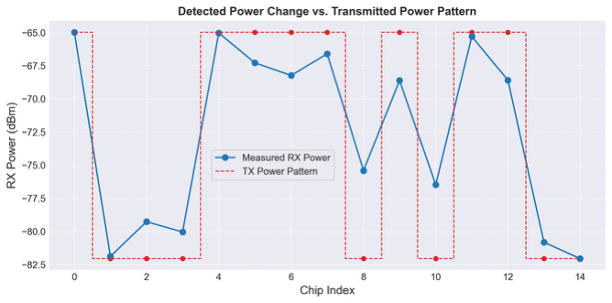}
    \caption{Transmitted pseudonym power pattern (red dashed) and measured received power pattern (blue solid) at OVRO/DSA.}
    \label{fig:power_patterns}
\end{figure}

\subsection{Why this watermark fits passive receivers}
This watermark design fits passive receivers because it does not depend on coherent demodulation. It only requires measuring power changes in time. Also, using a single subcarrier reduces receiver complexity: the observatory needs to extract only one frequency channel from the spectrogram. This supports lightweight coexistence monitoring and accountability without dedicated RF front-ends.

\section{Spectrogram-Only Pseudonym Receiver}
\subsection{Spectrogram-only processing}
The OVRO dataset includes spectrograms only and no IQ samples. Therefore, decoding is performed using energy-domain methods aligned with normal astronomy operation.

\subsection{Packet extraction by correlation}
We estimate the pseudonym packet start time by cross-correlation between the received power stream and a known transmitted reference packet pattern. The correlation peak gives an estimated start offset. This is robust because code modulation tolerates small synchronization error.

\subsection{Resolution mismatch and resampling}
A major challenge is mismatch between transmitter timing and receiver spectrometer timing:
\[
T_{\text{TX}} = \frac{1}{93.75~\text{kHz}} \approx 10.67~\mu\text{s},
\qquad
T_{\text{RX}} = \frac{1}{90~\text{kHz}} \approx 11.11~\mu\text{s}.
\]
Without correction, drift accumulates and the chip pattern is lost. We resample with ratio $25/24$ and oversample by factor 10.

One pseudonym bit is $N_{\text{TX}}=90$ samples. At the receiver:
\[
N_{\text{RX}} = \frac{90\cdot 90}{93.75} \approx 86.4.
\]
After resampling, each bit becomes 900 samples, giving around 60 samples per chip. We average power in chip intervals, which reduces noise and stabilizes decoding.

\subsection{Decoding and computation cost}
Decoding compares averaged chip-energy sequence to templates $\mathbf{c}_0$ and $\mathbf{c}_1$. The main computation is correlation and averaging. Since this uses only one frequency bin, the cost is low compared to full-band ML approaches \cite{Zhang2022deepRFI}. This suggests the receiver can be deployed as a lightweight post-processing step for accountability and reporting.

\begin{algorithm}[t]
\caption{Spectrogram-only pseudonym decoding}
\label{alg:spec_decode}
\begin{algorithmic}[1]
\REQUIRE Spectrogram $S(t,f)$, watermark bin $f_0$, templates $\mathbf{c}_0,\mathbf{c}_1$
\ENSURE Decoded bits $\hat{b}$
\STATE Extract $x[t]\leftarrow S(t,f_0)$
\STATE Start $\tau \leftarrow \arg\max \mathrm{corr}(x,\mathrm{template})$
\STATE Align $x[t] \leftarrow x[t+\tau]$
\STATE Resample $x \leftarrow \mathrm{Resample}(x,25/24)$
\STATE Average to chip energies $z[k]\leftarrow \mathrm{AvgChip}(x)$
\STATE Decode by matching $z$ with $\mathbf{c}_0,\mathbf{c}_1$
\end{algorithmic}
\end{algorithm}

\section{Experimental Setup}
\subsection{Transmitter}
A USRP B210 was used as the cooperative secondary transmitter. It was connected to a Bicolog 20300 broadband antenna (20~MHz--3~GHz). The transmitter operated at 920~MHz with sample rate 6~MHz. TX gain varied from 10~dB to 32~dB.

\subsection{Receiver}
The passive receiver is the OVRO/DSA-110 backend spectrometer. It provides 76.77~MHz observation bandwidth, divided into 853 channels with 90~kHz resolution each. The spectrometer time resolution is 11~$\mu$s.

\subsection{Dataset}
We collected around 70~GB of data. The dataset contains 20 files for 20 SNR conditions. Each file has matrix size $(460{,}000\times 853)$.

\begin{table}[t]
\centering
\caption{Field trial parameters.}
\label{tab:params}
\begin{tabular}{ll}
\toprule
\textbf{Parameter} & \textbf{Value} \\
\midrule
Center frequency & 920~MHz \\
OFDM sampling rate & 6 MHz \\
Subcarriers & 64 \\
Subcarrier spacing & 93.75 kHz \\
Watermark subcarriers & 1 \\
Backend bandwidth & 76.77 MHz \\
Frequency resolution & 90 kHz \\
Time resolution & 11 $\mu$s \\
SNR range & $-20$ to $+5$ dB \\
Dataset size & $\approx$70 GB \\
Conditions/files & 20 \\
\bottomrule
\end{tabular}
\end{table}

\begin{figure}[t]
    \centering
    \begin{subfigure}{0.24\textwidth}
        \centering
        \includegraphics[width=\linewidth]{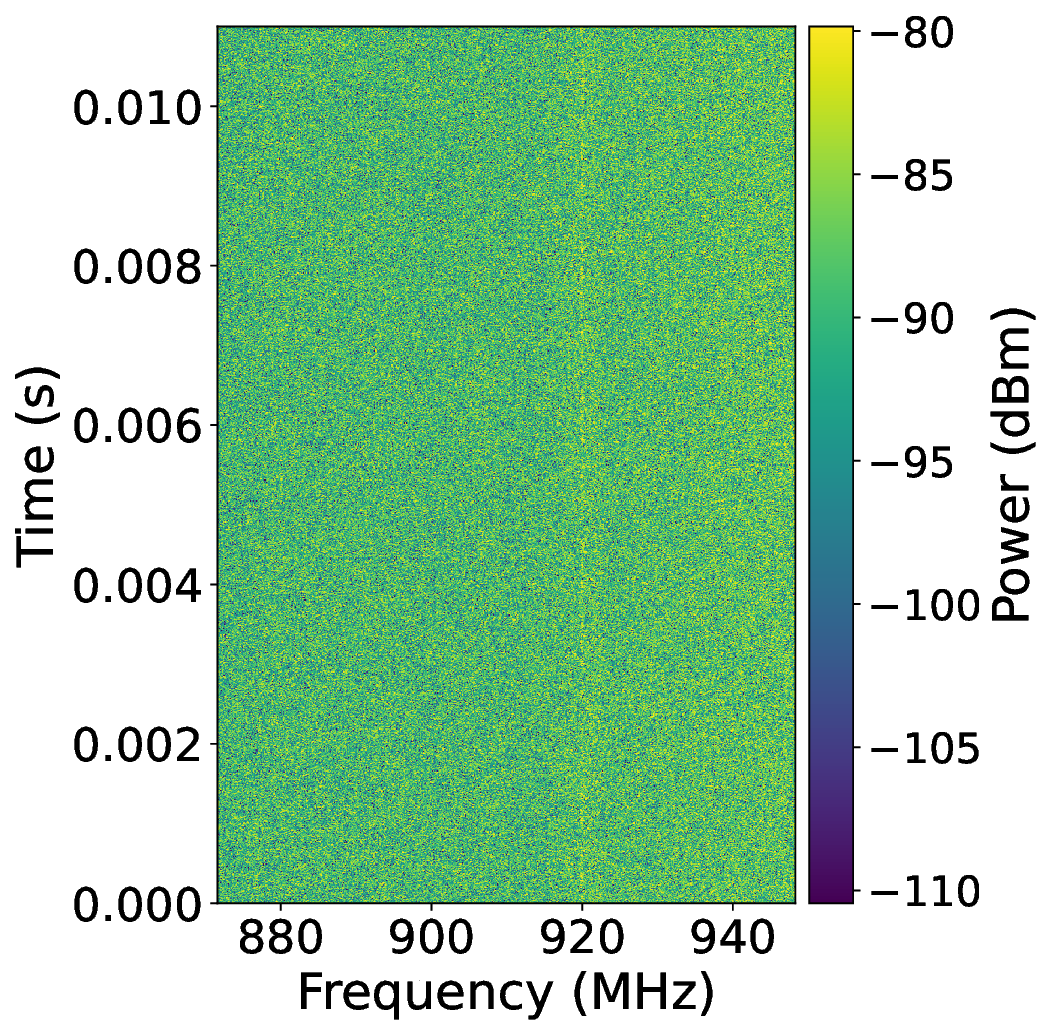}
        \caption{$-5$~dB}
        \label{fig:snr_-5}
    \end{subfigure}
    \hfill
    \begin{subfigure}{0.24\textwidth}
        \centering
        \includegraphics[width=\linewidth]{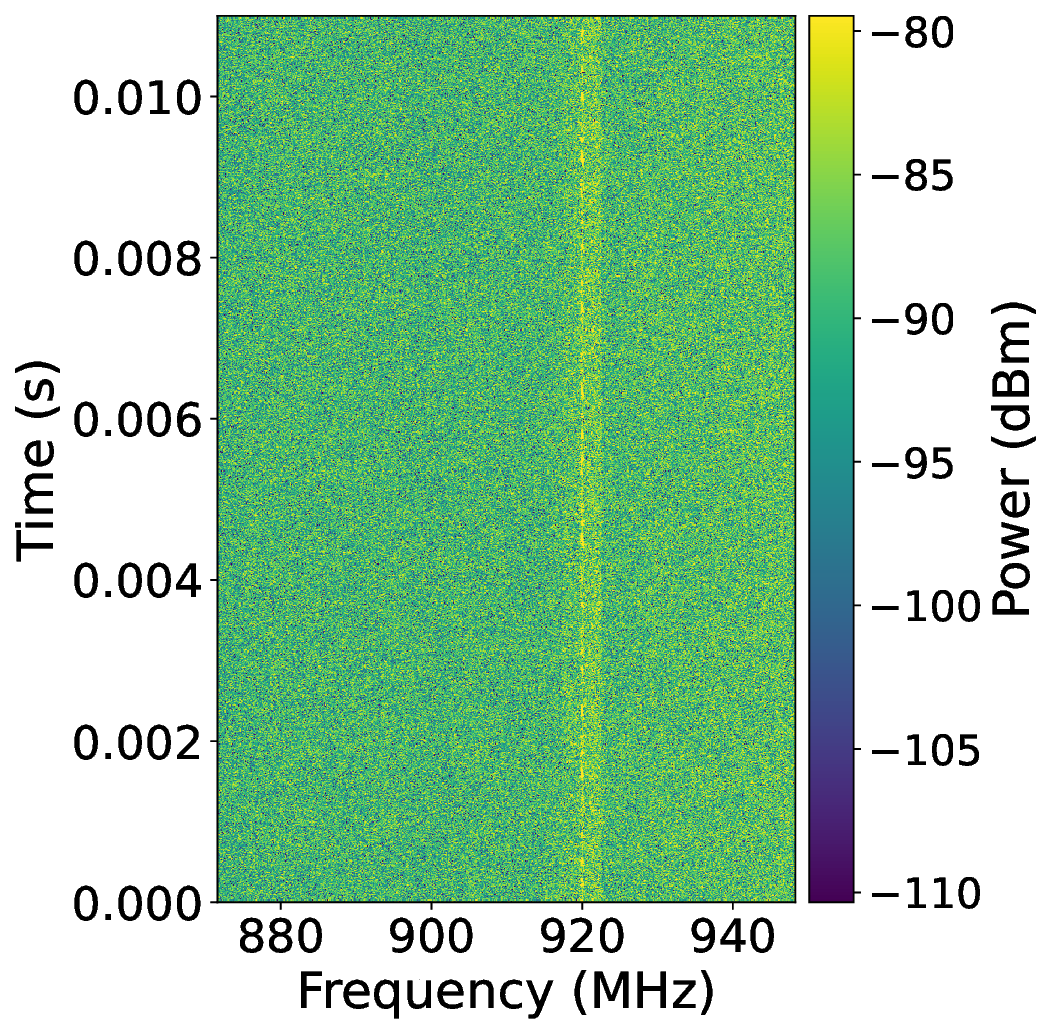}
        \caption{$+5$~dB}
        \label{fig:snr_5}
    \end{subfigure}
    \caption{Spectrogram examples at two SNR levels.}
    \label{fig:snr_spectrograms}
\end{figure}

\section{Results}
We decode pseudonym bits from spectrogram measurements and compute probability of pseudonym bit error $P_e$ versus SNR. To estimate SNR, we measure noise power by sampling the channel when the secondary transmitter is idle, and compute signal power from received packets during transmission.

\begin{figure}[t]
    \centering
    \includegraphics[width=\linewidth]{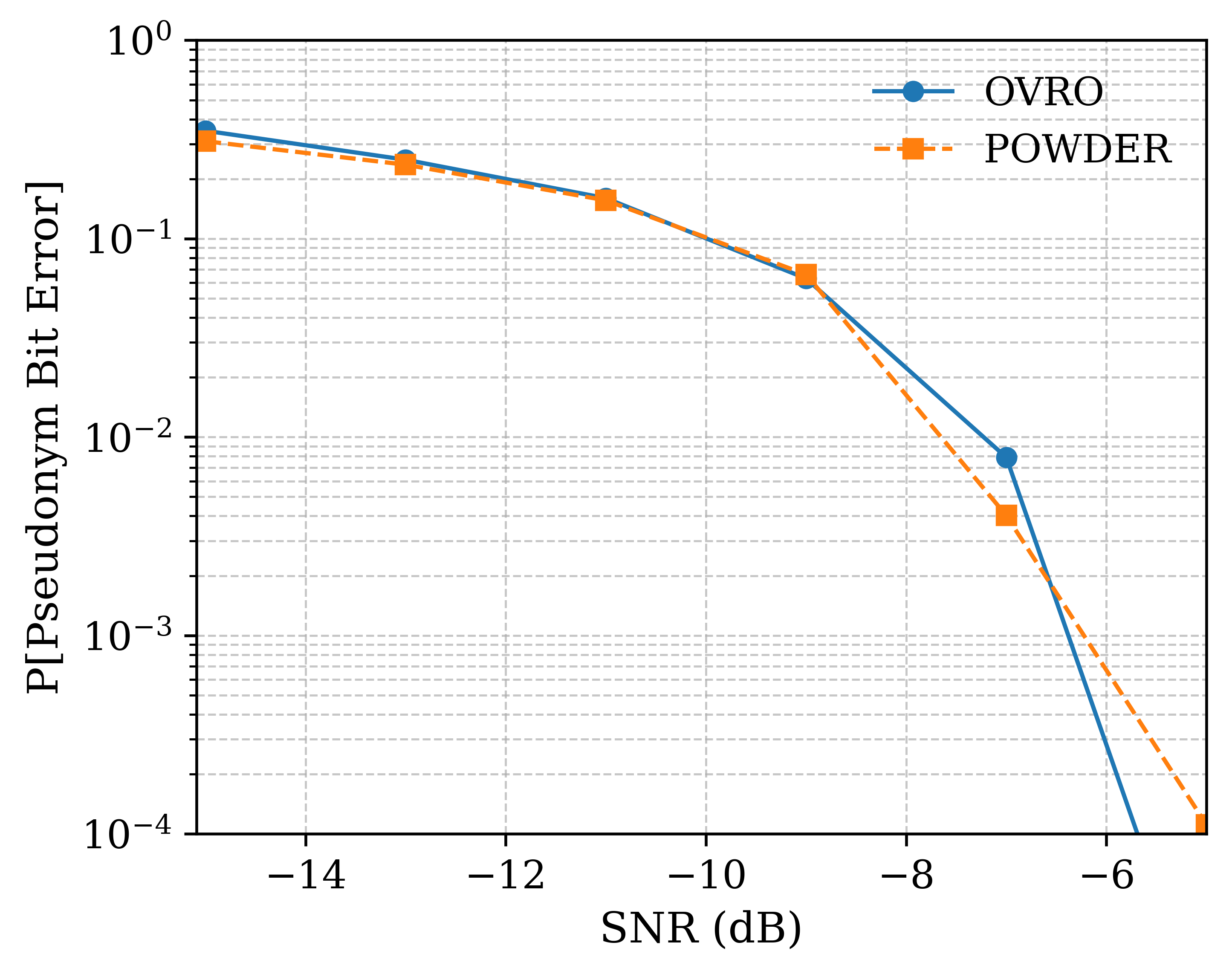}
    \caption{Probability of pseudonym bit error versus SNR for OVRO field trial.}
    \label{fig:prob_error}
\end{figure}

Figure~\ref{fig:prob_error} shows the bit error curve. The watermark can be decoded under low SNR. The error probability drops below $10^{-2}$ when SNR is higher than about $-8$~dB and approaches near error-free operation around $-6$~dB. This sharp transition happens because correlation peaks become stable and chip averages separate from noise.

At very low SNR ($<-10$~dB), correlation peaks become less stable and chip energy separation is weak, so decoding errors increase. Still, detection remains possible due to repetition across packets.

\textbf{Attribution behavior.} The purpose of Pseudonymetry is transmitter attribution rather than payload demodulation. Our field trial shows that pseudonym bits can be recovered from spectrogram-only observations in low-SNR regimes relevant to passive protection \cite{ITU-RA769-2}. This supports the feasibility of accountability-oriented coexistence where a cooperative transmitter can be linked to an interference event even when conventional demodulation is infeasible.

\textbf{Why this matters for coexistence.} If a passive receiver can attribute an interference event to a transmitter identity, it can support enforcement-oriented reporting and enable closed-loop interference management. This complements dynamic protection areas and adaptive protection zones \cite{Papadopoulos2023DPA,Munira2024DynamicPZ}, which require reliable monitoring and attribution to operate with reduced conservatism.

\section{Discussion and Limitations}
\textbf{Deployment relevance.} This work is relevant because it uses spectrogram-only data already produced by radio astronomy backends, reducing deployment cost and integration burden. It provides an accountability function that can complement classical mitigation pipelines \cite{Ellingson2005rfi,Ford2014rfi,Offringa2013} and learning-based RFI detection \cite{Zhang2022deepRFI}. Pseudonymetry is not intended to replace detection or classification; rather, it adds a \emph{coexistence-ready attribution channel} that supports operational enforcement and coordination in dynamic sharing frameworks.

\textbf{Baseline comparison.} Pure energy detection can indicate that RFI exists, but it cannot attribute the event to a transmitter. ML-based detection can help classify RFI patterns \cite{Zhang2022deepRFI}, but typically does not provide transmitter identity. Pseudonymetry adds identity under a cooperative transmitter assumption, enabling actionable reporting and cooperative shutoff for coexistence.

\textbf{Limitations.}
\begin{itemize}
    \item Cooperative transmitter assumption: attribution is not available for uncooperative emitters.
    \item Offline analysis: operational deployment requires real-time or near-real-time integration into monitoring pipelines.
    \item Spectrogram-only constraints: no phase/IQ, sensitive to distortion and offsets.
    \item Single interferer focus: multiple overlapping interferers and multiple simultaneous pseudonyms require further study.
\end{itemize}

\section{Conclusion}
We presented an OVRO field trial showing that pseudonym watermarks can be decoded from spectrogram-only backend data. Using an SDR-based OFDM transmitter at 920 MHz with a coded watermark on a single subcarrier, we designed a decoding pipeline that uses correlation-based synchronization and resampling-based mismatch correction. Results show robust low-SNR decoding using standard telescope backend products, supporting the feasibility of accountability-oriented coexistence near passive services. This provides a practical pathway to complement dynamic protection frameworks with attribution and enforcement primitives that enable less conservative spectrum sharing.

\bibliographystyle{ieeetr}
\bibliography{Ref} 
\end{document}